\documentclass[aps,prl,twocolumn,showpacs,superscriptaddress]{revtex4}
\usepackage{graphicx,color}

\begin{document}

\title{Relativistic Self-Induced Transparency Effect During Ultraintense Laser Interaction with Overdense Plasmas: Why It Occurs and Its Use for Ultrashort Electron Bunch Generation}

\author{V.I.~Eremin}
\author{A.V.~Korzhimanov}
\author{A.V.~Kim}
\affiliation{Institute of Applied Physics, Russian Academy of Sciences, 603950 Nizhny Novgorod, Russia}

\date{\today}

\begin{abstract}

A novel explanation of the relativistic self-induced transparency effect during superintense laser interaction with an overdense plasma is proposed. We studied it analytically and verified with direct modeling by both PIC and kinetic equation simulations. Based on this treatment, a method of ultrashort high-energy electron bunch generation with durations on a femtosecond time scale is also proposed and studied via numerical simulation. 
\end{abstract}


\maketitle

Recent progress in laser technology has paved the way of exploring previously unattainable regimes of ultraintense laser-plasma interaction \cite{bulanov}. In up-to-date experiments, laser intensities of the order of $10^{21}$ $\rm W/cm^2$ and higher can now be achieved, implying that the goals like the fast ignition concept for inertial confinement fusion (ICF) \cite{wilks}, laser-plasma based accelerators of charged  particles \cite{joshi,hegelich}, and compact sources of short-pulse x-ray radiation \cite{chen} might soon be within reach. It seems that this is exactly where fundamental physics and technological progress intersect. Here we address the fundamental issue of ultraintense electromagnetic (EM) wave  propagation through classically overdense plasmas, namely, why it occurs in the regime of relativistic self-induced transparency (SIT). In fact, this regime was first considered in the pioneer work \cite{akhiezer} in the form of stationary plane wave solution and later, in the 1970's, the stationary solutions were extended to non-homogeneous plasmas \cite{marburger,kaw}. Recently, another type of solution was found indicating that an ultraintense EM wave can penetrate into overdense plasmas over a finite length only, forming structured plasma distributions as a sequence of electron layers separated by about half a wavelength wide depleted regions, so that this strongly nonlinear plasma structure acts as a distributed Bragg reflector \cite{kim}. However, such analytical studies, done in a rather academic manner, do not provide an answer of how and why the propagation of a laser pulse into overdense plasma occurs. Numerical simulations of ultraintense laser interaction with overdense plasmas, on the one hand, have confirmed that the SIT effect takes place but also have revealed additional effects like anomalous longitudinal electron heating, which can change the optical properties of the plasma and strongly influence pulse propagation dynamics \cite{bonnaud,sakagami,guerin,tushentsov,berezhiani,ghizzo}. Nevertheless, the fundamental question of the SIT regime - why the penetration into classically overdense plasma in the form of traveling wave occurs – still has no reasonable explanation, especially taking into account strong peaking in the electron density in the front of the laser due to the ponderomotive force pushing electrons forward \cite{cattani}. It is generally recognized that the penetration occurs through lowering of the effective dielectric constant due to relativistic electron mass correction and plasma heating up to relativistic temperatures under the action of laser field. 

In this Letter, in order to get an insight into the physics we focus special attention on the boundary electrons dynamics and show that they are in unstable position, starting to be accelerated toward the incident wave. Thus, lowering of the effective plasma frequency occurs primarily due to the continuous generation of electron fluxes escaping into vacuum and thereby decreasing the electron density, which allows the laser pulse to propagate further into the plasma. The subsequent dynamics of the escaped electrons strongly depends on their interaction with a slowly moving (with velocities $\sim 0.1c$, where $c$ is the speed of light in vacuum) quasistanding wave structure comprising the incident and reflected EM waves. Eventually they are divided into two groups of electrons. The first one is coming back into the plasma forming an electron beam with sufficiently high density. Duration of this electron beam can be very short if an ultrashort pump laser is used. The second group of electrons is captured into potential walls and, in fact, it can be represented as two interpenetrating electron streams. Only these electrons gradually turn to heating. Indeed, the strong longitudinal heating takes place, especially for relatively high incident intensities, but it is rather concomitant effect than origin, although it can change even mode of the propagation. Based on the mechanism of electron beam generation we also propose here a new method of ultrashort electron bunch production on a femtosecond time scale.            

\begin{figure}
\includegraphics[width=0.4\textwidth,angle=0]{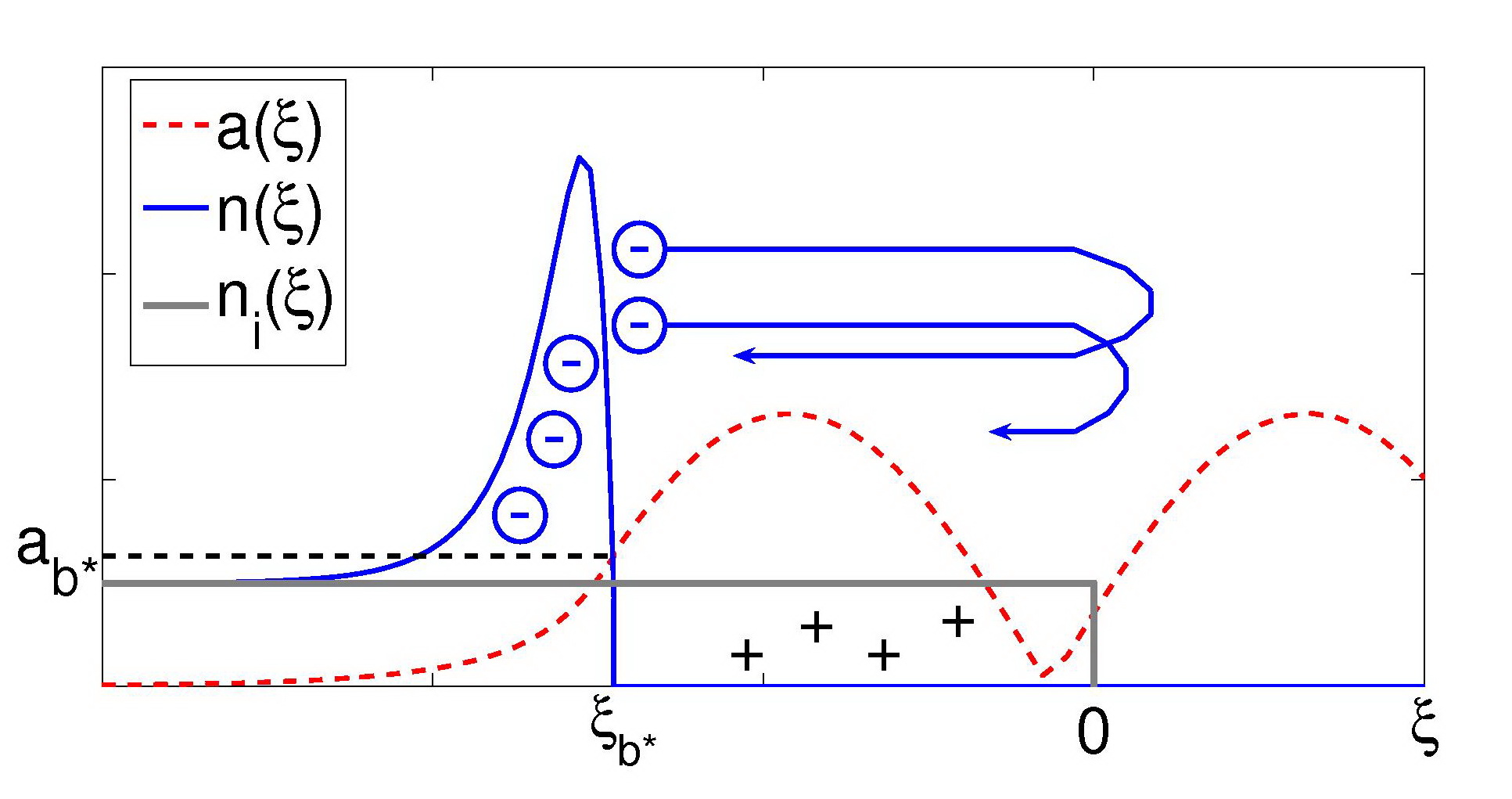}
\caption{Schematic of quasistationary model of laser-plasma interaction: spacial distributions of electromagnetic field $a$, electron $n$ and ion $n_i$ densities.} \label{ris1}
\end{figure}

We shall start our analysis with the relativistic cold fluid model and consider a circularly polarized normally incident EM wave from vacuum ($z>0$) onto a semi-infinite overdense plasma ($z\leq 0)$. In a linearly polarized wave, inherent ${\bf [j,B]}$ heating at relativistic intensities (see, e.g.,  \cite{lin}) essentially complicates the classical effect of SIT or even makes it significantly different from the latter. Since the relativistic SIT effect takes place on a time scale shorter than the ion response time, ion dynamics will be neglected. As was shown in \cite{marburger,kim,shen}, there are two different cases with respect to the plasma density $N_o$: (i) $N_o\leq 1.5N_c$ and (ii) $N_o > 1.5N_c$ ($N_c=m\omega^2/4\pi e^2$ is the critical density and $\omega$ is the laser frequency), which are also confirmed by computer simulations \cite{tushentsov,berezhiani}. In fact, at $N_o\leq 1.5N_c$, as was shown by Marburger and Trooper \cite{marburger}, in a stationary regime there is a continuous family of solutions for the everywhere positive electron density distribution, which means that penetration deep into plasma occurs in a "classical" way, through lowering the effective plasma frequency due to both relativistic and ponderomotive nonlinearities. Modeling the SIT effect using PIC simulations and the fluid approach \cite{tushentsov} showed quantitatively close results, indicating that kinetic effects are not important for this case, at least for incident intensities not much exceeding the threshold of penetration \cite{cattani}. The situation changes drastically at higher densities $N_o > 1.5N_c$, when the relativistic EM wave can considerably reduce the electron density in the region of its front until electron cavitation takes place. This electron cavitation acts as a "wall" making the plasma opaque to the propagation. In order to give an answer why the EM wave is able to propagate further deep into the plasma, we consider the dynamics of boundary electrons for the incident intensity near the threshold of the SIT regime. Our starting point is a solution given by Cattani {\em et al.}  \cite{cattani}, which can be depicted schematically as in Fig. \ref{ris1}.    
Let us obtain the electron density distribution from the Poisson equation using the balance of ponderomotive and electrostatic forces ($\phi^{\prime}=\gamma^{\prime}$):
\begin{equation}\label{eq:conc}
	n=1+\frac{\phi^{\prime\prime}}{n_o}=
		1+\frac{1}{n_o}\left(\frac{aa^{\prime\prime}}{\gamma}+\frac{a^{\prime2}}{\gamma^3}\right).
\end{equation}
Here we introduced normalized variables: $eA/mc^2=Re[a({\bf e}_x+i{\bf e}_y)e^{i\omega t}]$ is the vector potential; $\phi=e\varphi/mc^2$ the scalar potential; $n=N_e/N_o$ the electron density; $n_o=N_o/N_c$ the overdense parameter; $\gamma=\sqrt{1+a^2}$ the relativistic factor, the primes denote differentiation with respect to the longitudinal coordinate $\xi=\omega z/c$. By using the equation for vector potential, as in Ref. \cite{cattani}, for the incident intensity equal to the threshold value we arrive at the following equation for the field at the boundary where electrons vanish ($a(\xi_b=\xi_{b*})=a_{b*}$):
\begin{equation}\label{eq:thr_cond}
	2(n_o+a_{b*}^2)-3n_o\sqrt{1+a_{b*}^2}=0.
\end{equation}

Let us now show that the electrons at this boundary are in unstable equilibrium state. To do this, we take the ponderomotive force acting on them in the form
\begin{equation}\label{eq:pond}
	F_p=-\gamma^\prime=-\frac{aa^\prime}{\gamma},
\end{equation}
and calculate its gain when the electrons are shifted by a small distance $\Delta\xi$ opposite to the propagation direction, i.e., to the vacuum region ($\xi>\xi_b$).  Then for EM field we can use the vacuum solution, which represents a sum of incident and reflected waves, that should be self-consistently matched at the boundary with the field inside the plasma. Substituting this solution into Eq. (\ref{eq:pond}) and taking small variation from both parts, we obtain
\begin{equation}\label{eq:pond_incr_1}
	\Delta F_p=2\left(n_o-\frac{n_o+a_b^2}{\sqrt{1+a_b^2}}\right)\Delta\xi,
\end{equation}
where $a_b$ is the value of the field at the electron layer boundary.
For the electron to be in stable state when it is shifted from the equilibrium position, the total force acting on it must return it back to the equilibrium position. For that, the ponderomotive force gain must be more than the electrostatic force gain:
\begin{equation}
	\Delta F_p>-n_o\Delta\xi.
\end{equation}
From this we find the following condition on the field at the boundary:
\begin{equation}\label{eq:stab_cond}
	3n_o\sqrt{1+a_b^2}>2(n_o+a_b^2).
\end{equation}
Comparison of this expression with Eq.~(\ref{eq:thr_cond}) shows that the stability condition (\ref{eq:stab_cond}) ceases to be valid for the value of the field at $a_b=a_{b*}$, i.e., for the incident intensity equal to the threshold. This conclusion is highly important as it shows that all the skin-layer solutions are not only equilibrium (as follows from the balance of forces), but stable too. When the intensity of incident radiation exceeds the threshold, the electrons at the boundary become unstable and start to move toward the incident wave. This motion does not cease at least until electrons escape from the plasma region ($0>\xi>\xi_b$), which can be seen, for example, from Eq.~(\ref{eq:pond_incr_1}). This occurs because the gain in ponderomotive force only decreases with increasing field and, in spite of the fact that there may exist intervals where the field is less than at the electron layer boundary (near the standing wave nodes), this is insufficient for the balance of force. The balance is possible only when electrostatic force changes its sign, i.e., outside the plasma layer, $\xi>0$.

Direct calculation of ponderomotive force gain at the electron layer boundary with a small increase of incident intensity may be an additional argument confirming that the electrons start to move towards the laser, when the threshold is exceeded. Indeed, the substitution $a\rightarrow a+a_1$ into Eq. (\ref{eq:pond}), where $|a_1|<<|a|$, yields the following gain at the threshold intensity: 
\begin{equation}\label{pond}
	\Delta F_p=-\frac{n_oa_{b*}/\sqrt{1+a_{b*}^2}}{\left[2n_o(\sqrt{1+a_{b*}^2}-1)-a_{b*}^2\right]^{1/2}}.
\end{equation}
As seen from Eq. (\ref{pond}), the gain is negative, i.e., the ponderomotive force decreases and becomes less than the electrostatic force. Thus, with the excess over the threshold, the electrons start to escape from the plasma to vacuum and may get even far outside.  The maximal energy that may be acquired by the electrons is proportional to the magnitude of the charge separation field, which grows with increasing $n_o$, in particular, at $n_o\gg 1$ it can be estimated as $E\approx 1.3n_o^2$  \cite{cattani}. 

\begin{figure}
\includegraphics[width=0.55\textwidth]{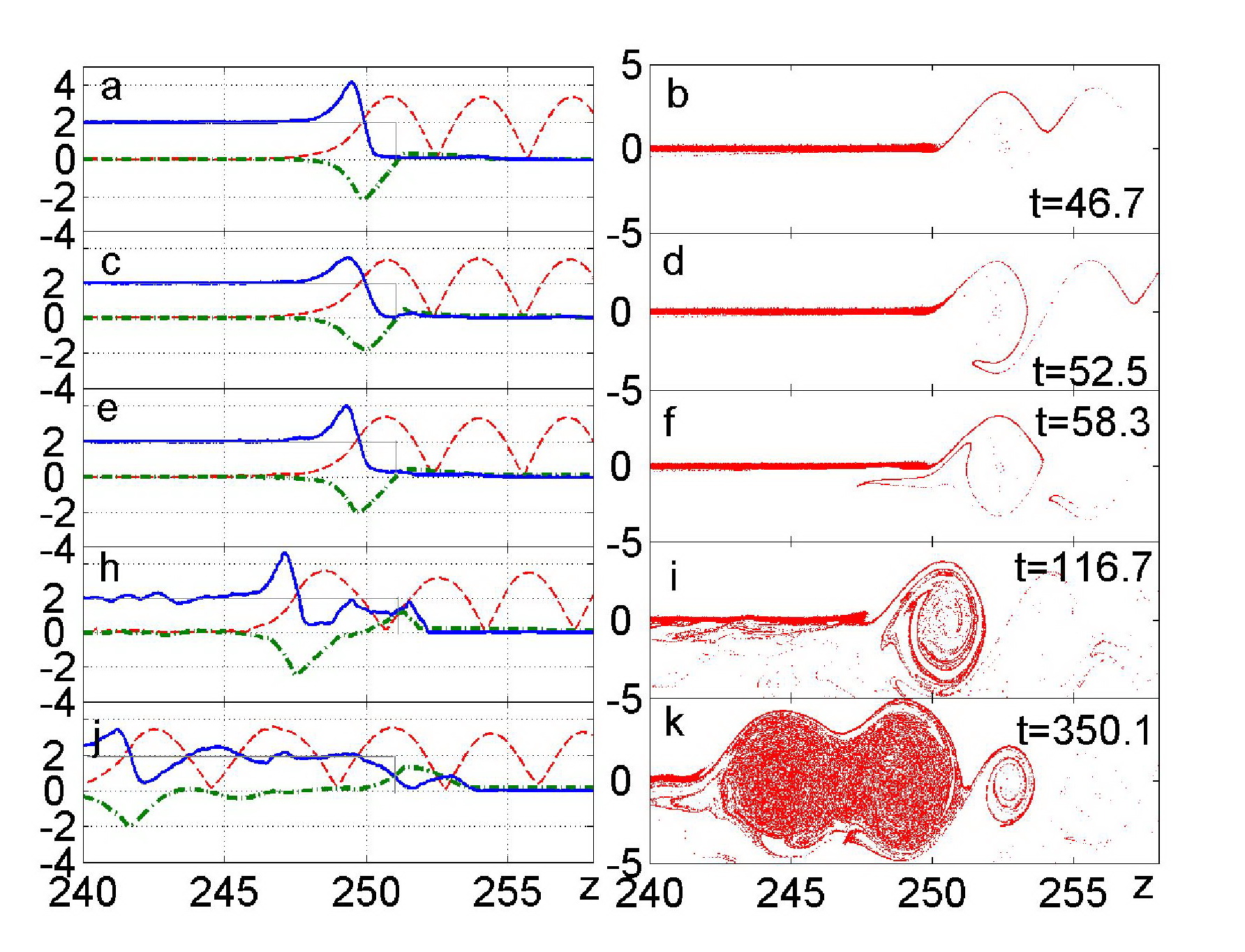}
\caption{Space-time dynamics (a,c,e,g,i) of the interaction of incident (on the right of the boundary) semi-bounded pulse of maximal amplitude $a_0=1.7$ with supercritical plasma $n_0=2$: distribution of electron density (continuous curve), ions  (dotted curve), absolute value of vector potential (dashed curve), scalar potential (dot-dash curve) at different moments of time. The corresponding phase portraits of the electrons are plotted on the right (b,d,f,h,j)} \label{ris2}
\end{figure}

\begin{figure}
\includegraphics[width=0.5\textwidth]{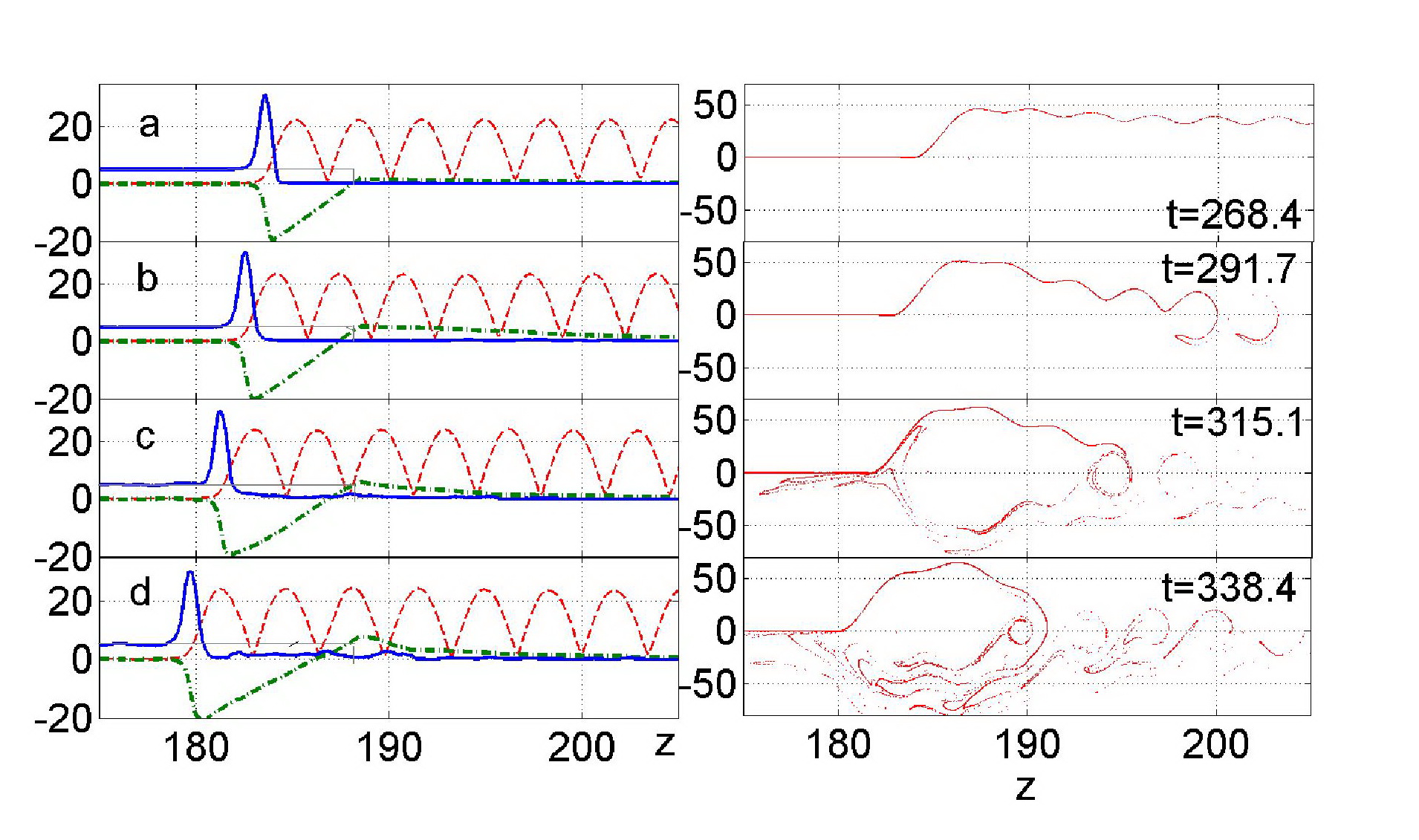}
\caption{The same as in Fig.~\ref{ris2} but for $a_0=12.8$ and $n_o=5$} \label{ris3}
\end{figure}

To consider full scenarios of the SIT propagation including kinetic effects, we have done a number of fully relativistic 1D PIC simulations for a wide range of plasma densities, $n_o=1-100$, paying special attention to the boundary electron dynamics. The incident laser pulse was taken either Gaussian or in the form of a unit function with smoothly growing amplitude up to maximal value exceeding the threshold. First, simulations showed good agreement with the SIT threshold values given in \cite{cattani}, which clearly indicates the validity of the cold fluid plasma model at intensities below the threshold. For these intensities the electron density at the vacuum-plasma boundary is redistributed under the action of ponderomotive force and a skin-layer structure is formed, as in Fig.~\ref{ris1}. They are similar to the field and electron density distributions of the dynamic problem presented in Figs.~\ref{ris2}(a),~\ref{ris3}(a), until the field at the boundary exceeds the threshold. In Figs.~2 and 3 the dynamics of laser propagation for two values of plasma density $n_o=2$ and $n_o=5$ is shown, respectively. Here analysis is given for a semi-bounded incident pulse with rising time equal to 3 laser periods (time and space variable are normalized as: $t\rightarrow \omega t$, $z\rightarrow \omega z/c$) and with maximal amplitude ($a_{0}=1.7$ and 12.8) slightly exceeding the threshold  ($a_{th}=1.58$ and 12.6, respectively). At the beginning, simulations show that the dynamics of the interaction is quasi-stationary, i.e. the electromagnetic field structure having the form of a quasi-standing wave with a skin-layer distribution in plasma gradually pushes electrons deep inside the medium, creating an uncompensated ion charge at the plasma boundary and thus increasing the longitudinal quasistatic field. When the incident intensity becomes higher than the threshold, violation of the quasi-stationary interaction starts, in the first place, from the acceleration of boundary electrons in the skin layer towards the incident wave, which is readily traced in the phase portrait of longitudinal motions (see Figs.~\ref{ris2}(b,d,f,h,j) and \ref{ris3}(b,d,f,h)) and is in qualitative agreement with the above analytical considerations. The difference between Fig.~\ref{ris2} and Fig.~\ref{ris3} is that for higher plasma density the electrons acquire higher energies and may propagate at large distances from plasma into vacuum. As the electrons escape, the plasma eventually acquires a positive charge, thus the electrons get into the total field that is a sum of the quasi-static field and the standing wave which can be regarded as a sequence of potential wells formed by the ponderomotive potential. With a small excess over the threshold, the electrons both oscillate, primarily in the first potential well, and return to plasma forming electron beams moving deep into plasma as is shown in Figs.~\ref{ris2}(f,h,j) and \ref{ris3}(f,h). It should be noted that subsequent dynamics of the escaped electrons is quite complex and strongly depends on laser and plasma parameters. However, eventually there are two groups of electrons. The first one is coming back into the plasma in the form of an electron beam that can propagate deep into the plasma and has sufficiently high density. Moreover, the duration of this electron beam may be very short if an ultrashort pump laser pulse is used. The second group of electrons is captured into potential walls and, in fact, it may be represented as two interpenetrative electron streams, clearly indicating that the hydrodynamic approach is violated. Only these electrons gradually turn to heating. This is clearly seen on the right-hand side of Figs.~\ref{ris2} and \ref{ris3}, where the phase trajectories inside the separatrices corresponding to potential wells are gradually densely filled everywhere, which is the evidence of effective electron heating up to the temperature corresponding to the ponderomotive potential.
With a substantial excess over the threshold, the electrons may move over large areas of a standing wave. Note that the presence of oscillatory trajectories for continuously incoming electrons in ponderomotive potential is indicative of the need to use the kinetic description because the electron fluxes moving in forward and backward directions intersect here. It is also worth noting that the electron beam moving deep into plasma has a quite high density, only an order of magnitude less than the initial plasma density and energy of several MeV; parameters of the beams corresponding to Figs.~\ref{ris2} and \ref{ris3} are the following: density $n_b\simeq0.2$ and 1.3, maximal kinetic energy (in units of the rest energy $mc^2$) $\epsilon_{bmax}\simeq0.63$ and 19.3 with energy spread of $\Delta\epsilon_b=0.4$ and 1.3, respectively.

The process of acceleration of electrons from plasma to vacuum and then back is evidently accompanied by simultaneous wave penetration deeper into plasma. In this case, the wave reflects from a plane gradually moving into a plasma that separates the transparent and nontransparent regions. The penetration velocity of the incident wave, or to be more exact of the reflection plane, depends on the excess over the threshold of SIT. If the threshold is exceeded substantially, the velocity is $v \approx 0.1-0.2 c$, which agrees with data in \cite{bonnaud,guerin,ghizzo}. However, for a small excess over the threshold, after penetration to a finite depth, the interaction enters a quasi-stationary stage close to that described in the hydrodynamic approximation  \cite{tushentsov}, except for the absence of purely cavitation regions. These regions are partially filled by electrons,  which indicates at least that for high local ponderomotive potential the effective electron temperature is high. 

\begin{figure}
\includegraphics[width=0.55\textwidth]{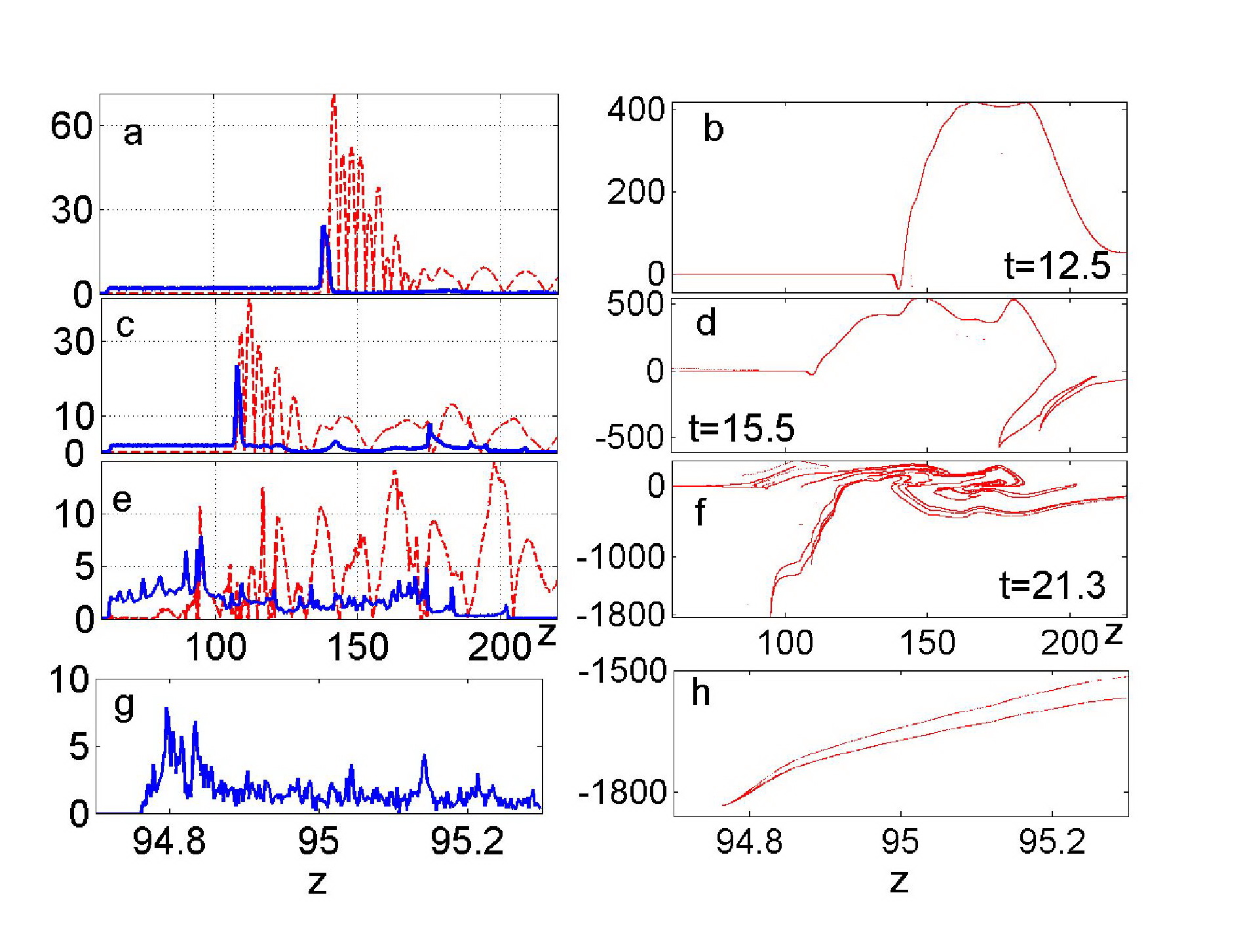}
\caption{The same as in Fig.~\ref{ris2}, but for the incident 5-fs Gaussian pulse with $a_0=60$. Data for the generated electron beam: (g) snapshot of the density distribution and (h) its detailed phaze portrait} \label{ris5}
\end{figure}

In view of the above result of electron beam acceleration, we propose here a new method of high-energy electron beam generation with ultrashort durations. The first advantage of this method is that the density of the beam can be very high, for example, for the parameters as in Fig.~\ref{ris2} it is of the order of $0.1n_o$, just ten times less than the solid density but with energy of several MeV. On the other hand, by using superintense ultrashort laser pulses, electron bunches with a few femtosecond or even shorter duration may be produced. The interaction of a 5-fs laser pulse of intensity $I\simeq10^{22}$Wcm$^{-2}$ ($a_0=60$) with plasma of $n_o=2$ is depicted in Fig.~\ref{ris5}, where an electron bunch generated at the plasma boundary has a sub-fs duration, energy of order GeV and maximal beam density of $10^{22}$ cm$^{-3}$ ($n_b\sim 10$). Results of simulations for various incident intensities are systematized in Table~1 where parameters of the generated electron beams are listed: $n_{bmax}$ -- maximal electron density, $<n_{b}>$ -- beam density averaged over the time interval indicated in the last column, the same for the kinetic energy of the beam $\epsilon_b$. Two averaging times are used because the temporal structure of the generated beams exhibits quite a narrow peak and a relatively extended pedestal, as is shown in Fig.~\ref{ris5}(g); it allows us to present results on electron bunch structure in a better way.

Table 1.\\
\begin{tabular}{|c|c|c|c|c|c|c|}
\hline $n_0$ & $a_0$ & $n_{bmax}$ & $<n_b>$ & $\epsilon_{bmax}$ & $<\epsilon_b>$ & $<t>$\\
\hline 2 & 1.7 & 0.23     &	0.16  & 1.69     &	1.28      & 0.6\\
\hline 2 & 1.7 & 0.23     &	0.08  & 1.69     &	0.94      & 5.1\\
\hline 2 & 3.1 & 1.8      & 0.44  & 11.3     &  3.6       & 0.6\\
\hline 2 & 3.1 & 1.8      & 0.32  & 11.3     &  2.2       & 2.7\\
\hline 2 & 4.7 & 9.56     &	2.98  & 18.8     &	6.7       & 0.6\\
\hline 2 & 4.7 & 9.56     &	1.25  & 18.8     &	2.7       & 2.4\\
\hline 2 & 7.5 & 12.09    & 4.5   & 58.7     &	47.8      & 0.6\\
\hline 2 & 7.5 & 12.09    & 2.22  & 58.7     &  26.9      & 3.6\\
\hline 2 & 30  & 11.06    &	9.9   & 700      &	684       & 0.7\\
\hline 2 & 30  & 11.06    &	1.82  & 700      &  369.1     & 6.0\\
\hline 2 & 60  & 14.89    &	6.39  & 1809     &	1658      & 0.6\\
\hline 2 & 60  & 14.89    &	1.93  & 1809     &	1321      & 6.3\\
\hline
\end{tabular}	

In conclusion, we have presented a new treatment of the relativistic self-induced transparency effect that allows ultraintense laser pulses to penetrate deep into overdense plasmas. This treatment is based on the careful consideration of the boundary electrons dynamics, which for incident intensities exceeding the threshold start to move towards the laser, thus decreasing the electron density in front of the pulse. The subsequent dynamics of the generated electron beams strongly depends on the laser and plasma parameters, in particular, they may be considered as ultrashort high-energy electron bunches. In view of such electron bunch dynamics, we also propose here a method of high-energy beam generation in a femtosecond time scale durations or even shorter.

\end{document}